\documentclass[runningheads]{llncs}

\usepackage{times}

\usepackage[utf8]{inputenc}
\usepackage[small]{caption}
\usepackage{booktabs}
\usepackage{enumitem}
\usepackage{soul}
\usepackage{url}
\usepackage{amsmath,amssymb,amsfonts}
\usepackage{algorithmic}
\usepackage{graphicx}
\usepackage{textcomp}
\usepackage{xcolor}
\usepackage{mathrsfs}
\usepackage{multirow}
\usepackage{bbm}
\usepackage{xspace}
\usepackage{subcaption}
\usepackage{xcolor,colortbl}
\usepackage[ruled,vlined,linesnumbered]{algorithm2e}
\newcolumntype{a}{>{\columncolor{pink!15}}c}
\newcolumntype{q}[1]{>{\centering}p{#1}}

\newcommand{\srcomment}[1]{#1}

\newcommand{\comment}[1]{}

\newcommand{\xhdr}[1]{\vspace{0.3mm}\noindent{{\bf #1.}}}
\newcommand{\systemname}{\textsc{CoLAB}\xspace}

\begin{document}	
	\title{Modeling Implicit Communities using Spatio-Temporal Point Processes \\ from Geo-tagged Event Traces}
	
		\author{Ankita Likhyani\inst{1} \and
				Vinayak Gupta\inst{2}\and
				Srijith PK \inst{3} \and
			    Deepak P \inst{4}\and
		        Srikanta Bedathur \inst{2}}
			%
			%
			\institute{Indraprastha Institute of Information Technology, Delhi \and Indian Institute of Technology, Delhi \and Indian Institute of Technology, Hyderabad \and Queen's University, Belfast \\ \email{ankital@iiitd.ac.in, } \email{Vinayak.Gupta@cse.iitd.ac.in, } \email{srijith@iith.ac.in, } \email{deepaksp@acm.org, and } \email{srikanta@cse.iitd.ac.in}}
		
	
	\maketitle
	
	\begin{abstract}
	The location check-ins of users through various location-based services such as Foursquare, Twitter and Facebook Places, etc., generate large traces of geo-tagged events. These event-traces often manifest in hidden (possibly overlapping) communities of users with similar interests. Inferring these implicit communities is crucial for forming user profiles for improvements in recommendation and prediction tasks. Given only time-stamped geo-tagged traces of users, can we find out these implicit communities, and characteristics of the underlying influence network? Can we use this network to improve the next location prediction task?
	In this paper, we focus on the problem of community detection as well as capturing the underlying diffusion process and propose a model \systemname based on spatio-temporal point processes in continuous time but discrete space of locations that simultaneously models the implicit communities of users based on their check-in activities, without making use of their social network connections. \systemname captures the semantic features of the location, user-to-user influence along with spatial and temporal preferences of users. To learn the latent community of users and model parameters, we propose an algorithm based on stochastic variational inference. To the best of our knowledge, this is the first attempt at jointly modeling the diffusion process with activity-driven implicit communities. We demonstrate \systemname achieves upto 27\% improvements in location prediction task over recent deep point-process based methods on geo-tagged event traces collected from Foursquare check-ins. 
\end{abstract}
		
	\section{Introduction}
	Proliferation of smartphone usage and pervasive data connectivity have made it possible to collect enormous amounts of mobility information of users with relative ease. Foursquare announced in early 2018 that it collects more than $3$ billion events every month from its $25$ million users\footnote{https://bit.ly/2BdhnnP (accessed in February 2019)}. These events generate a location information diffusion process through an underlying --possibly hidden-- network of users that determines an \emph{location adoption} behavior among users. Location adoption primarily depends upon user's spatial, temporal and categorical preferences. For instance, one user's check-in at a newly opened jazz club could inspire another user to visit the same club or a similar club in her vicinity depending upon the distance from the club and time of the day/week. These users might not be having a social connection but it's an implicit influence because of similar choices. This often leads to the formation of --possibly overlapping-- communities of users with similar behavior. Detecting community of such like minded people from large geo tagged events can benefit applications of various domains such as targted advertisements and friend recommendation. Prior work \cite{10.1145/2492517.2492576} also suggests that social connections are not as effective factors for prediction tasks. 
	
	In this paper, we move away from communities derived purely from social network links, and instead explicitly identify spatio-temporal activity-driven communties. Eschewing the reliance on social connections alone allows us to learn communities that are not overly biased towards connected users. At the same time, it supports the use of \systemname under settings where social network is not available -- either due to privacy settings of users, or due to other restrictions from platform. 
	
\begin{figure}
	\centering
	\begin{subfigure}[b]{0.33\textwidth}
		\includegraphics[width=\linewidth, height =4cm]{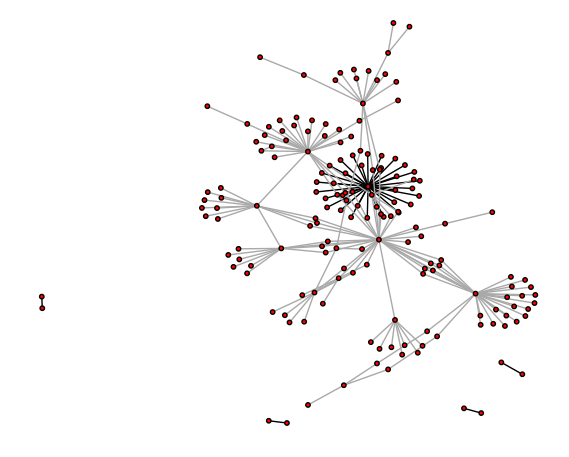}
		\caption{threshold = 0.5}
	\end{subfigure}%
	\begin{subfigure}[b]{0.33\textwidth}
		\includegraphics[width=\linewidth, height =4cm]{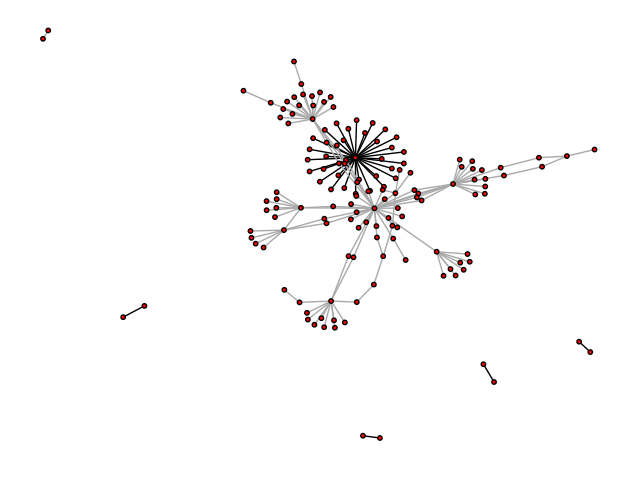}
		\caption{threshold = 0.7}
	\end{subfigure}%
	\begin{subfigure}[b]{0.33\textwidth}
		\includegraphics[width=\linewidth, height =4cm]{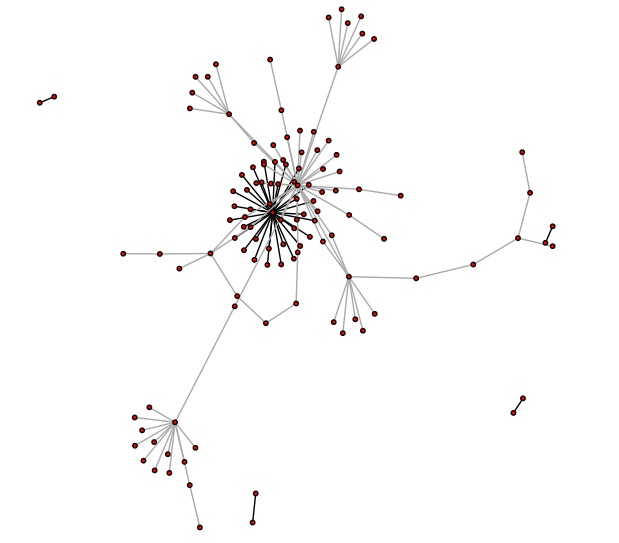}
		\caption{threshold = 0.9}
	\end{subfigure}  
	\caption{\systemname extracts underlying diffusion network over US data using geotagged checkin traces. This Maximum Weighted Spanning Forest (MWSF) is constructed by varying the threshold for edge weights (i.e. influence score computed using \systemname). The inferred influence network depicts a tree-like structure of influence.}
	\label{fig:comm}
	
\end{figure}

	\begin{example} 
		We illustrate the spatio-temporal activity-based network and communities that \systemname derives using the US dataset we collected, shown in Figure \ref{fig:comm}. We observe that explicit social connections, shown using black edges, are very sparse and are unable to capture the implicit influence network. On the other hand, the latent network derived by  \systemname, shown using gray edges, can identify significantly higher number of relations in the latent influence network. It can be observed that it is not only captures the clusters but also identify potential influencers (and their influence networks) which can critically help in prediction tasks.
	\end{example}

	In this paper we determine that the location adoption process and \emph{community formation} among users can be explained by the same latent factors underlying the observed behavior of users without considering their social network information.
%
	We propose \textbf{\systemname} (Communities of Location Adoption Behaviour) that focuses on jointly inferring location specific influence scores between users and the communities they belong to, based solely on their activity traces. \textbf{\systemname} completely disregards the social network information, which makes it suitable for scenarios where only activity traces are available. Note that if social network is available, it can be used as a prior or a regulization over the influence matrix we derive. Unlike the current best-performing models for location prediction task (e.g.,~\cite{zarezade-2018}), \systemname avoids the community formation from being biased only by the availability of social connections. Thus we generate better communities even for users with few or no social connections as shown in figure \ref{fig:comm}. 
	
Further, we generate \emph{overlapping communities} of users that take into account the spatio-temporal patterns, and the shared special interests over location categories. Although a user may be part of multiple communities with different special interests, we assume each event or check-in to be associated with only one of those communities. Prior works have focused on modeling temporal + textual together \cite{10.1145/2492517.2492576} or temporal + spatial features together \cite{doi:10.1137/18M1226993}, none of the techniques to the best of our knowledge have modeled the entire combination of temporal, spatial and location semantics in a spatio-temporal point process to infer the underlying influence network.
	

\xhdr{Contributions} Our main contributions are as follows:
\begin{enumerate}
	\item We propose a novel model called \systemname to model activity patterns over geo-tagged event traces. It leverages \emph{spatio-temporal Hawkes process}\cite{spatialhawkes14,sthp} to not only construct a information diffusion-based latent network but also recover overlapping community structures within it.
	\item We develop a novel \emph{stochastic variational inference technique} to learn the latent communities and model parameters.
	\item As our target is to identify communities that comprise users who share interests as opposed to just being socially connected, there is unfortunately no gold-standard community information to evaluate our results. Therefore, we first empirically evaluate our method over synthetic data; further, results shows our inference algorithm can accurately recover the model parameters. For communties evaluation on real data, we make use of a joint loss function that evaluates on the basis intra-community properties defined in terms of users' category affinity and spatial dispersment in their checkin characteristics.
	\item We evaluate on two real-world geo-tagged event traces collected from two countries -- viz., SA (Saudi Arabia) and US. The experimental results demonstrates that we achieve upto 27\% improvement over neural network based models.
\end{enumerate}

	
	\section{Related Work}\label{sec:relwork}

In this section, we briefly review existing literature on community detection and characterization of location adoption. 

\subsection{Community Detection}

Within general social networks, there has been much work on detecting communities as overlapping or non-overlapping clusters of users such that there is a high degree of connectedness between them. Techniques have largely considered social-network connectedness as the main driver in forming communities. Community detection techniques could adopt a top-down approach starting from the entire graph and form communities by separating them into more coherent subsets that would eventually become communities. Methods from this school include those that use graph-based methods~\cite{prat2014high} and filtering out edges based on local informations such as number of mutual friends~\cite{zhao2012large}. Analogously, community discovery could proceed bottom-up by aggregating proximal nodes to form communities. Techniques within this category have explored methods such as merging proximal cliques~\cite{kumpula2008sequential} or by grouping nodes based on affinities to 'leader' nodes~\cite{khorasgani2010top}. Point processes\cite{MR1950431}, which are popular models for sequential and temporal data, have been recently explored for community detection in general social networks\cite{DBLP:conf/sdm/TranFSZ15}. NetCodec, the method proposed therein, targets to simultaneously detect community structure and network infectivity among individuals from a trace of their activities. The key idea is to leverage multi-dimensional Hawkes process in order to model the relationship between network infectivity vis-a-vis community memberships and user popularity, in order to address the community discovery and network infectivity estimation tasks simultaneously. Our task of community detection within LBSNs is understandably more complex due to the primacy of spatial information in determining LBSN community structures. Accordingly, we leverage a spatio-temporal variant of multi-dimensional Hawkes process, in devising our solution.

\noindent{{\bf LBSN Community Detection:}} There has been existing work ~\cite{DBLP:conf/icwsm/NoulasSMP11a,10.1007/978-3-642-35386-4_9,7876373,LI2018188} on community detection in LBSNs that leverage features such as venue categories, temporal features, geo-span, social-status, structure based and location based features in determining community structure within clustering formulations. They have used standard ML based techniques such as Spectral clustering \cite{DBLP:conf/icwsm/NoulasSMP11a}, $M^2$\cite{10.1007/978-3-642-35386-4_9}: a k-means based clustering, Entropy-based \cite{7876373}, and Sequence-Mining \cite{LI2018188} based techniques. Our method, as illustrated therein, models a wider variety of features, incorporating both spatio-temporal check-in information as well as venue category information, in inferring communities and user-user influence. 

	
	
\subsection{Location Adoption Characterization}

With each LBSN check-in being associated with a location, the location information is central to LBSNs. There has been much research into modelling user-location correlations in various forms, which may be referred to as {\it location adoption} in general. 

\cite{DBLP:conf/sdm/JankowiakG17} is the closest work to our work, where authors determine the patterns from geo located posts from twitter. Our model deals with discrete locations unlike \cite{DBLP:conf/sdm/JankowiakG17} which simply considers sampling of spatial locations from a continuous distribution which can not be used for modeling check-in locations. The intensity function in CoLAB seamlessly integrates temporal and spatial components of the model and jointly learns the parameters whereas in \cite{DBLP:conf/sdm/JankowiakG17} location and time are modeled separately. Moreover, we use multi-dimensional Hawkes Process as intensities differs across users. The parameter estimation of multi-dimensional Hawkes Process is much harder and Monte Carlo Simulations based methods used in \cite{DBLP:conf/sdm/JankowiakG17} are very slow. We introduce the use of stochastic variational inference (SVI) \cite{blei17} to overcome this. \cite{conf/icwsm/GaoTL12} and \cite{Gao:2013:MTE:2505515.2505616} exploit the role of social correlations and temporal cyclical behaviors in improving upon the task of predicting the next location that a user would check-in. Location categories (e.g., restaurant, pub etc.) has also been seen to be useful for next location prediction\cite{Likhyani:2015}. \cite{Ye:2011:EGI:2009916.2009962} exploit geographical influence across locations in recommending POIs (points of interest) to LBSN users. 

Another task that has attracted significant attention within LBSNs is that of quantifying user influence in LBSNs. \cite{Zhang2012} consider using geo-social correlation between users to estimate mutual influence. \cite{Wu2013} leverage the observation that mutual influence is central to successful recommendation to identify a set of influential seed nodes using information cascade models. \cite{Bouros:2014} adopt influence propagation models to address a slight variant, that of identifying region-specific influential users. \cite{Li2014} and \cite{Wang:2016} propose other kinds of models for the influence maximization task within LBSNs. ocation Promotion \cite{Zhu2015}\cite{ijcai2017-314} and Trip Purpose Detection \cite{HU2017136} form other les addressed tasks within the context of LBSNs. Mobility models that mine spatial patterns based on generative models \cite{conf/icwsm/GaoTL12}, Gaussian distributions \cite{Cho:2011:FMU:2020408.2020579} and kernel density based estimations \cite{Lichman:2014:MHL:2623330.2623681} have been particularly popular in modeling location adoption behavior within LBSNs.

	\section{Problem Statement}
	
Consider an geo-tagged event trace dataset $S$ over a set of $L$ locations $\mathbb{L} = \{ \ell_k \}_{k=1}^L =  \{ (x_k, y_k) \}_{k=1}^L$,  a set of $I$ users $\mathbb{U} = \{ i_k\}_{k=1}^I$ and $V$ categories (restaurants, entertainments etc.). Let us consider there are $N$ events, with the $n^{th}$ check-in denoted as $E_n = ( t_n, \ell_n, c_n, i_n, g_n)$ and $\mathbb{E} = \{ E_n \}_{n=1}^N$. The notation denotes that $E_n$ is the check-in event involving the user $i_n$ checking in to location $\ell_n$ at time $t_n$, with the category associated with the location being $c_n$ and the \emph{latent community} of the user associated with the check-in is $g_n$. \srcomment{The task is to learn the latent community associated with the users and effectively model the diffusion of information among users.  Towards this, we aim to  learn a matrix $\phi$ of size $|\mathbb{U}|$ x $|M|$ where $i^{th}$ row represents community participation for the $i^{th}$ user (assuming $M$ communities). In addition, to model the diffusion process, we estimate matrix $A_{ij}$, where an element $a_{ij}$ represents the influence of $i^{th}$ user on $j^{th}$ user. }

\section{Preliminaries}
	
	\begin{table}
		\begin{tabular}{l p{0.75\columnwidth}}
			\toprule
			\textbf{Symbol} & \textbf{Description} \\
			\midrule
			$\mathbb{U}$ & set of $I$ users \\
			$\mathbb{L}$ & set of $L$ locations \\
			$\mathbb{G}$ & set of $M$ communities (latent) \\
			$\lambda_i (t, x, y)$ & intensity at time $t$ and spatial coordinate $x,y$ of user $i$ \\
			$\mu_i$ & base rate of checkins of user $i$ \\
			$A_{ij}$ & influence of user $i$ on user $j$ \\
			$\kappa(t, x, y)$ & triggering / self exciting kernel \\
			\bottomrule
		\end{tabular}
		\caption{Terminology}
		\label{tab:terminology}
	\end{table}

\xhdr{Hawkes Process~\cite{hawkes-oakes-1974}} It is a point process with self triggering property, has been explored to model temporal occurrences of events in social media~\cite{zhao15}. The conditional intensity is modelled as:
$\lambda(t) = \mu + \sum_{t_k < t} \kappa (t - t_k),$
where $\mu$ is the base intensity of event occurrence, with $\kappa()$ (kernel function) modelling the influence of past events using an exponential decay function. 


\xhdr{Multidimensional Hawkes Process} This extension allows modelling of influence of events across entities (e.g., users in the social media case) by considering time-stamped events from across entities~\cite{pmlr-v28-zhou13}; this naturally yields to modelling infectivity across users. The intensity function for an entity $i$ at time $t$ depends on past events as:
$\lambda_i(t) = \mu_i + \sum_{t_k < t} A_{i_ki} \kappa(t - t_k), $
where $(t_k, i_k)$ represents historical user/entity events on time scale, $\mu_i$ $>$ 0 being the base intensity for the $i^{th}$ entity/user, the non-negative influence matrix $A_{ij}$ quantifying the quantum of influence from user/entity $i$ to $j$. 
	
	

\xhdr{Spatio-Temporal Hawkes Process:} This extends the basic Hawkes process along the spatial dimension\cite{spatialhawkes14}. These simultaneously model spatio-temporal clustering behavior and has been found useful for modelling dynamics of epidemics and crime. The intensity function takes the form:
\[ \lambda(\ell,t | H_t) = \mu + \sum_{t_k < t} \kappa(\ell - \ell_k, t - t_k), \]
where $\ell_k$ is the location associated with the event at $t_k$. 


	\section{CoLAB Model}
	\label{sect:model}
	\srcomment{In this section we describe our model to infer the $\phi$ matrix i.e. the communities vector inferred from check-in activities and information diffusion over  the users in the network.}	
	
	\subsection{Spatio-Temporal Data Modeling}
		Given the check-in events $E$ as defined earlier, for modeling time and spatial components w.r.t. communities, we define the Hawkes process based model as follows:
	
	\subsubsection{Intensity Function}  

%
%

We model the user's community-specific intensity using the multi-dimensional spatio-temporal Hawkes process. Multi-dimensional because influence from other users also contribute in the intensity of a user~\cite{pmlr-v28-zhou13}. Consider the task of estimating the \emph{community-specific} intensity of a user $i_n$ towards generating a check-in $E_n = (t_n, \ell_n, c_n, i_n, g_n)$; the multi-dimensional spatio-temporal Hawkes process formulation yields the following:

\begin{equation}
\lambda_{i_n, g_n} (t_n,\ell_n) =  \mu_{i_n}\eta_{g_n} + \sum_{t_k < t_n} A_{i_ki_n} \kappa (t_n - t_k, \ell_n-\ell_k) \mathbb{I}(g_k = g_n)
\end{equation}

where $\mu_{i_n}$ is the base intensity of user $i_n$ and $\eta_{g_n}$ is weight associated to $g_n$ towards a community $g_n$, and $\boldsymbol{\eta} = \{\eta_g | g = 1, ..., M \}$ with $\eta_{g} \geq 0$. $\lambda_{i_n,g_n} (t_n, \ell_n)$ is community specific intensity of user $i$ at the $n^{th}$ instance. We allow historical check-ins to contribute to the intensity - proportionate to their temporal and spatial proximity to $t_n$ and $\ell_n$ respectively, and weighted using the influence between user $i_n$ and $i_k$ (i.e., $A_{i_ki_n}$) - as long as they belong to the same community, enforced by the indicator function $\mathbb{I}(g_k = g_n)$. Here, $\kappa (t_n - t_k, \ell_n - \ell_k)$ is the triggering exponential kernel which factorises over time and location.
	\begin{equation}
	\kappa (t_n - t_k, \ell_n - \ell_k) = \kappa(t_n - t_k) * \kappa (\ell_n - \ell_k) ,
	\end{equation}
   where, $\kappa(t_n - t_k) = \exp (- \nu (t_n - t_k))$ is the time specific triggering kernel with $\nu$ decay and $\kappa(\ell_n - \ell_k) = \frac{1}{2 \pi\, h} exp \left(-\frac{  || \ell_n - \ell_k ||} {2 h} \right)$ is the location specific triggering kernel with $h$ bandwidth.  When the decay parameter is low, the influence of the previous events is high and similarly when the bandwidth parameter is high the influence of previous locations is high. 

In general,  the intensity of a particular user $i$ at some time $t$ and location $\ell$, is given as the sum of intensities that are estimated at the level of each community i.e. total intensity $\lambda_{i} (t, \ell)$  = $\sum_ g \lambda_{i, g} (t, \ell)$

	\subsection{Category Distribution}
	\label{cat_dis}
	\srcomment {The category $c$ associated with a  check-in is represented as a $|V| $-length vector and it represents one  of $V$ possible categories\footnote{We overload the notation $c$ to  also represent a scalar categorical value in the set $\{ 1,\ldots, V\}$} such as restaurant, entertainment etc. associated with the check-in.  Also, we assume that the category depends on the underlying latent community associated with this check-in. For example, some community may be more inclined towards restaurants while another community is oriented towards sports. The category  is modelled as a sample from a Multinomial (categorical) distribution,}
	\begin{equation}
	c \sim Multinomial(\boldsymbol{\theta_{g}})
	\end{equation}
where $\boldsymbol{\theta_{g}}$ is a $| V| $-length vector whose elements encode probability of each category and which depends on the community $g$ that the check-in belongs to. \srcomment{We assume a prior over $\boldsymbol{\theta_{g}}$ as a sample from Dirichlet distribution with parameters $\boldsymbol{\theta_0}$. We write the conditional distribution $p(c, \boldsymbol{\theta_g}|\boldsymbol{\theta_0})$  as:}
	\begin{equation}
	p(c, \boldsymbol{\theta_g}|\boldsymbol{\theta_0}) = p(c| \boldsymbol{\theta_g}) p(\boldsymbol{\theta_g} | \boldsymbol{\theta_0}) =
	\theta_{g,c} \frac{\Gamma (\sum_{j} \theta_{0,j})}{\prod_{j} \Gamma(\theta_{0,j})} \prod_{j} \theta_{g,j}^{\theta_{0,j}-1}
	\end{equation}
	$j$ runs over the $V$ categories and $p(c| \theta_g)$ is given as $\theta_{g,c}$
	\subsection{Distribution over communities}	
	We assume the latent variable $g$ (the communities) associated with the user for some check-in, is distributed as multinomial distribution parameterized by $\pi_i$ for a user $i$. $\pi_{ig}$ represents the probability user $i$ belongs to community $g$. 
	\begin{equation}
	g \sim  Multinomial(\boldsymbol{\pi_i})
	\end{equation}
	\subsection{Generative Process}


\begin{algorithm}
	\small
	\SetAlgoLined
	Initialize the number of communities $M$, and number of checkins $N_i$ for each user\;
	Set $\mu_i$ proptional to $N_i$\;
	Initialize $A_{ij}$ as column normalized matrix\;
	Initialize $\pi$, $\eta$ and $\theta$ as Dirichlet-Multinomial distribution \;
	Initialize $\lambda_i(t_0, x_0, y_0) = \mu_i  \quad \forall i=1,\ldots, U$ \;
	\For{n = 1 to N}{
	    Sample ($t_n, \ell_n$) from $\sum_{i=1}^{U} \lambda_i (t,\ell)$\;
	    Sample $i_n$ from Multinomial ($\lambda_1(t_n,\ell_n)$, $\lambda_2(t_n,\ell_n)$ , ... , $\lambda_U(t_n,\ell_n)$)\;
	 	Sample $g_n$ from a Multinomial ($\pi_{i_n}$)\;
		Sample $c_{n}$ from $Multinomial(\theta_{g_n})$ ($\theta_g$ is defined in section \ref{cat_dis})\;
	}
	\caption{Generative Process}
	\label{sec:gen_pr}
\end{algorithm}

	
	 Note that, for sampling (t,x,y), the thinning algorithm proposed in \cite{doi:10.1002/nav.3800260304} is modified in order to sample location coordinates from discrete ``venue'' set rather the continuous space. 
	First we consider a discrete set of locations $L$ for the user based on her region. We sample $(x', y')$ at $n^{th}$ iteration, from a Gaussian distribution centered at the previous coordinates in the $(n-1)^{th}$ iteration: $(x_{n-1}, y_{n-1})$. Once $(x',y')$ is sampled, the nearest coordinate in the $L$ is determined and returned as $(x_n,y_n)$.
	
\section{Estimation and Inference}
Given the multi-dimensional Hawkes process model defined above, the joint probability density function over the check-in events $\mathbb{E}$ is given as:


\begin{equation}
\prod_{n=1}^N p(t_n, l_n, c_n, g_n| i_n) = \prod_{n=1}^N  \biggl ( (p( t_n, l_n | i_n, g_n) \times p(g_n|i_n) ) \times p(c_n | g_n, \theta) \biggr )
\end{equation}

Here, $p(c_n | g_n, \theta) =  \theta_{g_n,c_n}$, where $c_n$ represents the category associated with the  $n^{th}$ check-in, and $p(g_n|i_n) =  \pi_{i_n,  g_n}$ the probability that user $i_n$ belong to the community $g_n$. 
\comment{
\begin{eqnarray}
	\prod_{n=1}^N p(t_n, \ell_n | i_n, g_n) = \prod_{n=1}^N \lambda_{i_n,g_n}(t_n, \ell_n) ( 1 - F^*(t_n,\ell_n)) 
\end{eqnarray}
where, $F^*(t_n, \ell_n)$ is the cumulative distribution function. and $( 1 - F^*(t_n,\ell_n)) $ correspond to the probability of no new events before time $t_n$ ~\cite{MR1950431}. Thus, 
}
	\begin{equation}
	\prod_{n=1}^N p(t_n, \ell_n | i_n, g_n) =  \prod_{n=1}^{N} \lambda_{i_n,g_n}(t_n, \ell_n) \exp \biggl (-\sum_{i=1}^{U} \int\limits_0^{T} \int\limits_{\ell_{min}}^{\ell_{max}} \lambda_{i}(t, \ell) dt d\ell \biggr) 
	\label{eqn:likelihood}
	\end{equation}
is the likelihood (event density) of generating the observations given the community and users in the interval $[(0, \ell_{min}),(T, \ell_{max})]$.  The first term in (\ref{eqn:likelihood}) provides the 
instantaneous probability of occurrence of the observed events and the second term provides the probability that no event happens outside these observations (survival probability)~\cite{MR1950431}.   
	\comment{
and 	
	\begin{equation}
	\prod_{n=1}^{N}  p(c_n | g_n, \theta) = \prod_{n=1}^{N} \theta_{g_n,c_n}
	\end{equation}	
Here, in this section we consider $c_n$ is a scalar value representing the category associated with the  $n^{th}$ check-in.	
The product over the categories is given as 	
	\begin{equation}
	\prod_{n=1}^{N} p(g_n|i_n) = \prod_{n=1}^{N} \boldsymbol{\pi_{i_n g_n}}
	\end{equation}
Thus the complete joint likelihood assuming the community is observed can be written as 
	\begin{multline}
	L =  \prod_{n=1}^{N} \bigg[ \bigg(\lambda_{i_n}(t_n, x_n, y_n)   \times \pi_{{i_n,g_n}} \bigg) \times \theta_{g_n,c_n} \bigg] \\ \exp \bigg(-\sum_{i=1}^{U} \int\limits_0^{T} \int\limits_{\ell_{min}}^{\ell_{max}} \lambda_i(t,\ell) dt d\ell \bigg)
	\end{multline}
	}
Thus, the complete joint log likelihood is: 
	\begin{multline}
	\label{eqn:ll}
	\mathcal{LL} =  \sum_{n=1}^{N}  \bigg( \log \lambda_{i_n,g_n}(t_n, \ell_n) + \log \pi_{i_n, g_n}  + \log \theta_{g_n,c_n} \bigg) \\ - \sum_{i=1}^{U} \int\limits_0^{T} \int\limits_{\ell_{min}}^{\ell_{max}}  \lambda_{i}(t, \ell) dt d\ell
	\end{multline}
Assuming communities are known, we can estimate the model parameters $\mu$, $\eta$,$A$, $\theta_{g}$'s and $\pi$'s by maximum likelihood estimation. We treat the kernel parameters, and the Dirichlet parameters as the hyper-parameters which we initialize to some fixed values. However, the communities are latent and the maximum likelihood estimation cannot be applied directly. This calls for the expectation maximization algorithm, where the parameters are estimated after integrating out the latent variables from the joint likelihood using the posterior distribution over the latent variables. In our case, the posterior distribution over latent communities is given as
\begin{multline}
 p(g_1, \ldots, g_n | \{ t_n,\ell_n, c_n, i_n\}_{n=1}^N)  \\ 
 = \frac{\prod_{n=1}^N p( t_n, \ell_n | i_n, g_n)  \times p(c_n | g_n, \theta) \times p(g_n|i_n) }{\sum_{g_1, \ldots, g_n} \prod_{n=1}^N p( t_n, \ell_n | i_n, g_n)  \times p(c_n | g_n, \theta) \times p(g_n|i_n) } 
 \label{eqn:posterior}
\end{multline}
The posterior distribution over the latent communities cannot be obtained in closed form due to the intractable normalization constant (denominator term) which involves an exponential number of summation terms. 
Markov chain Monte Carlo methods~\cite{bishop06} can be used to obtain samples from the posterior. However, these approaches are not scalable to large datasets~\cite{blei17} and becomes computationally expensive for use in LBSNs.  To overcome this, we use a  variational expectation maximization  algorithm where we approximate the posterior over communities using a variational distribution and estimate the model parameters and variational parameters by maximizing a variational lower bound~\cite{avi17}.

\subsection{Variational Expectation Maximization}
	
\comment{	
		 $\mu, A_{ij}$, and $\pi$ are the set of parameters to be estimated, $X = <t, l, c, u>$ is the set of observed events and $G$ is the set of latent communities to be estimated
		
		The likelihood function of the complete data $<t, l, c, u>$ $=$ $\{<t_n, l_n, c_n, g_n>\}$ where $t_0$ = 0, $x_0$ = $x_{min}$ and $y_0$ = $y_{min}$, and $t_N$ = T, $x_N$ = $x_{max}$ and $y_N$ = $y_{max}$ is given as		
		\begin{multline}
		L(t,l,c,g,\theta) =  \prod_{n=1}^{N} \bigg[ \bigg(\lambda_{i_n}(t_n, \ell_n)   \times \boldsymbol{\pi_{i_n g_n}} \bigg) \times \theta_{g_n,c_j}^{c_j} \bigg] \\ \exp \bigg(- \sum_{i=1}^{U} \int\limits_0^{T} \int\limits_{\ell_{min}}^{\ell_{max}}  \lambda_i(t,\ell) dt d\ell \bigg)
		\end{multline}
}	
		 The latent variables  $g_n$'s dependent on different types of feature set i.e. space, time through $p(t_n, \ell_n | i_n, g_n) $ and semantics  through $p(c_n | g_n, \theta)$. Though the prior over $g_n$ is conjugate to  $p(c_n | g_n, \theta)$, it is not with respect to  $p(t_n, \ell_n | i_n, g_n) $ and hence the posterior over $g_n$ cannot be computed in closed form.  Moreover, $g_n$'s are inter-dependent i.e. at current step $g_n$ it is dependent on history from $g_1$ to $g_{n-1}$ as well as the future ones i.e. $g_{n+1}$. Thus marginalizing out over such interconnected latent variables to compute the normalization constant for the posterior is intractable. To this end we assume a  variational distribution over $g_n$'s  conditioned on the user $i_n$. The conditional variational distribution over $g_n$ is considered to be a multinomial distribution  with parameters $\boldsymbol{\phi_{i_n}}$. The variational parameter $\boldsymbol{\phi_{i}}$ for a user $i$ represents the posterior probability distributions over the communities for the user as observed from the data. 	
		\begin{equation}
		\label{eqn:VD}
		q({g_n}| i_n) =   Multinomial(g_n | \boldsymbol{\phi_{i_n}})
		\end{equation}		
		\comment{
		 We next lower bound log likelihood $\mathcal{L}(t, \ell , c, g)$ with respect to $q$
		\begin{multline}
		\mathcal{L}(t, \ell, c, g) = \log \sum_{g} L(t, \ell, c, g) \\
		= \log \sum_{g} \frac{L(t, \ell, c, g)}{q(g)} q(g) \\	
		= \log E_q \bigg(\frac{L(t, \ell, c, g)}{q(g)} \bigg)
		\end{multline}
		Using Jensen's Inequality for concave functions (i.e. $\log$) we have :
\begin{equation}
		\mathcal{L}(t, \ell , c, g) \geq E_q ( \log L(t, \ell, c, g)) - E_q (\log q(g) ) \equiv \mathscr{L}
\end{equation}
		 The right hand side $\mathscr{L}$ is the Evidence Lower Bound (ELBO), which will be used as the surrogate to the true log-likelihood in inference and learning. Thus we have,
}		

 The variational parameters  can be learnt by minimizing the KL divergence between the variational posterior (\ref{eqn:VD})  and the exact posterior (\ref{eqn:posterior}). However, a direct minimization of KL divergence is not possible due to the intractable posterior.  Following  variational inference approach~\cite{blei17},  the variational parameters are learnt  by maximizing  a variational lower bound, Evidence Lower Bound (ELBO),   which indirectly minimizes the KL divergence. ELBO is obtained by considering an expected value of the  complete joint log likelihood w.r.t the variational distribution~\cite{blei17} and acts as a lower bound to the marginal likelihood or evidence (normalization constant of the posterior). Hence, ELBO is useful to learn the model parameters also  in addition to the variational parameters. 
Using the  variational distribution defined in (\ref{eqn:VD}) and the complete joint log likelihood (\ref{eqn:ll}), we obtain the ELBO as:
		\begin{multline}
		\label{eqn:ELBO}
		\mathscr{L} =   \sum_{n=1}^{N} \bigg( \mathbb{E}_q[\log \lambda_{i_n,g_n}(t_n, \ell_n)] + \sum_{m=1}^{M} \phi_{i_n, m} \log \pi_{i_n, m}  +  \sum_{m=1}^{M}  \phi_{i_n, m} \log \theta_{m,c_n}   \bigg)  \\ -\sum_{i=1}^{U} \int\limits_0^{T} \int\limits_{\ell_{min}}^{\ell_{max}}  \mathbb{E}_q [\lambda_{i}(t, \ell)] d\ell dt - \mathbb{E}_q [ \log q ]
		\end{multline}
Here, $\mathbb{E}_q$ represents the expectation with respect to the variational distribution $q$ defined in (\ref{eqn:VD}).  We learn the variational parameters and the model parameters by maximizing the ELBO.  Table~\ref{tab:parameters} lists the model parameters and variational parameters to be learnt using ELBO.    All the terms in the ELBO except the first term can be computed in closed form. 

	\begin{table}[t]
        \centering
	\caption{Parameters to be estimated and whether a Hyperparameter}
		\label{tab:parameters}
	\begin{tabular}{ c  p{6.5cm}  c }	
		\toprule
		\textbf{Par} & \textbf{Description} & \textbf{\emph{H}} \\
		\midrule
		$\boldsymbol{\mu}$ & Base Intensity &  \\
		$\boldsymbol{\eta}$ & Weight associated towards community & \\
		$A_{ij}$ & Influence Matrix & \\
		$h$ & Bandwidth (KDE) & $\checkmark$\\
		$\nu$ & Temporal Decay Parameter & $\checkmark$ \\
		$\boldsymbol{\theta_0}$ & Dirichlet Prior: Category & $\checkmark$ \\ 
		$\boldsymbol{\theta_{g}}$ & Multinomial Prior: Categories / community & \\	
		$\pi $ & Multinomial Prior: Communities (All users) &  \\ 
		$\phi$ & Variational Parameters: Communities (All users) &  \\
		\bottomrule
	\end{tabular}
	\end{table}	
		Since the first term in (\ref{eqn:ELBO}) cannot be computed in closed form, we approximate it using the samples from the variational posterior (\ref{eqn:VD}) (Monte-Carlo approximation). 		
		\begin{equation}
		\mathbb{E}_q[\log \lambda_{i_n,g_n}(t_n, \ell_n)]   \approx \frac{1}{S} \sum_{s = 1}^{S} \log \lambda_{i_n,\boldsymbol{g^{(s)}}} (t_n,\ell_n)
		\end{equation}		
		where $\boldsymbol{g^{(s)}}$  represent the vector of $N$ samples  sampled from the joint variational distribution over all the $g_n$'s, \textit{i.e.} $q(\boldsymbol{g^{(s)}}) = \prod_{i=1}^N q(g_n^{(s)} | \phi_{i_n})$.  This results in a stochastic variational lower bound where the stochasticity arises due to the approximation of expectation using Monte Carlo sampling~\cite{avi17}. 
We learn the model parameters $\mu$, $\eta$, $A_{ij}$, $\theta$ by maximizing the stochastic variational lower bound~\cite{conf/icml/PaisleyBJ12}  using gradient based methods. However  learning the variational parameters is problematic as the variational parameters does not appear explicitly in the stochastic term but only through the samples.  For determining gradient w.r.t. $\phi$ we apply the Reinforcement trick to the stochastic term and compute the gradient as follows~\cite{gal16}:		
		\begin{equation}
\bigtriangledown_\phi \mathbb{E}_q[\log \lambda_{i_n,g_n}(t_n,\ell_n)] \\ \approx \frac{1}{S} \sum_{s=1}^{S} \log \lambda_{i_n,\boldsymbol{g^{(s)}}} (t_n, \ell_n) \bigtriangledown_{\phi} \log q(\boldsymbol{g^{(s)}})          
\end{equation}

\section{Experiments}
\textbf{Compared Baselines} We empirically evaluate the performance of \systemname\footnote{We pledge to make our codes and datasets public.} over synthetic and real datasets with the following baselines: 
\begin{itemize}[leftmargin=*]
\item \textbf{STHP}: Spatio-Temporal Hawkes Process models the diffusion process across spatial and temporal dimensions but ignores the \textit{category} associated location feature. This is the baseline derived from \systemname ignoring the categories. 
\item \textbf{Sequence Mining} \cite{LI2018188}: First extracts frequent occuring venue category sequences and assigns communties based on clusters with similar patterns. 
\item \textbf{DH} \cite{dh}: Dirichlet-Hawkes, clusters continuous time event streams using a modified Hawkes model with preferential cluster assignment through Dirichlet Process.
\item \textbf{RMTPP}:\cite{10.1145/2939672.2939875} Recurrent Marked Temporal Point Process model the time and the marker information by learning a general representation of the nonlinear dependency over the history based on recurrent neural networks. In this model, event history is embedded into a compact vector representation which is then used for predicting the next event time and marker type.
\end{itemize}
For an even comparison, we feed the check-in events to all the baselines and evaluate the community quality formed by DH and Sequence Mining along with location prediction for STHP and RMTPP.

\subsection{Synthetic Data} 
We generate synthetic data using algorithm \ref{sec:gen_pr} (statistics in Table~\ref{tab:syn}). The set of locations, \#users to categories i.e. (U:V) and to the number of checkins i.e. (U:N) are kept similar to the real data collected from Brazil with $\nu$ (temporal decay parameter) is set to 0.01 \cite{Yang:2013} and $h$ is picked up from the bandwidth values learned from users' checkins. During inference, we use \emph{true values} for all the parameters, except for the influence matrix $A_{ij}$ and the user-community posterior $\phi$ which are estimated. We use $RelErr(A_{ij},\hat{A}_{ij}) = \frac{1}{I^2}\sum_{i,j = 1}^{I}  \frac{| a_{ij} - \hat{a}_{ij} |} {| a_{ij} |} $ (and similarly for $\phi$), as the metric to evaluate the ability to recover the true values. Table~\ref{tab:relerr} indicates that the stochastic variational inference technique offers considerably better reconstruction of the parameters, recording significant reductions in the error.

\begin{table}%
	\begin{minipage}[t]{\columnwidth}
		\centering
		\caption{Dataset Properties}%
		\label{tab:syn}%
		\begin{tabular}{l@{\hspace{0.5cm}}llll}
			\toprule
			\textbf{\footnotesize Property} & 	{\footnotesize\bf\#Users(U)} & {\footnotesize\bf\#Communities(M)} & {\footnotesize\bf\#Categories(V)} & {\footnotesize\bf\#Check-ins(N)} \\
			\midrule
			\textbf{Synthetic} &  100 & 10 & 200 & 9777\\ 
			\textbf{SA (Real)} & 95 & - & 314 & 15110 \\
			\textbf{US (Real)} & 133 & - & 524 & 22059 \\
			\bottomrule
		\end{tabular}

	\end{minipage}
	\hfill
	\begin{minipage}[t]{\columnwidth}
		\centering
		\caption{$RelErr$ on $A$ and $\phi$ and True Positives for Location Prediction results at Top-K on Synthetic Data}%
		\label{tab:relerr}%
			\begin{tabular}{l c@{\hspace{0.5cm}} c@{\hspace{0.5cm}}  c@{\hspace{0.5cm}} c}
				\toprule
				& \multicolumn{2}{c }{RelErr} & \multicolumn{2}{c}{Top K}  \\
				\cmidrule(lr){2-3}\cmidrule(lr){4-5} 
				\textbf{Technique} & $A$ & $\phi$ & 5 & 10 \\
				\midrule
				STHP & 0.99174 & 0.13807 & 681 & 1206 \\
				\systemname & \textbf{0.04813} & \textbf{0.07216} &  \textbf{972} &  \textbf{1677} \\
				\bottomrule
			\end{tabular}

	\end{minipage}
\end{table}

%


\subsection{Real Data}
For real data, we use our crawls over Foursquare conducted between January-2015 and March-2016 and construct two collections consisting of check-ins from Saudi Arabia (SA) and United States (US), with details given in Table~\ref{tab:syn}. We allocated first  80\% (as per check-in timestamp) of each dataset to \textit{training} and remaining for \textit{testing}. Here, we also use temporal decay parameter as 0.01 \cite{Yang:2013} and $h$ is learnt for each user based on Silverman’s rule of thumb in kernel density estimation \cite{bishop06} and is fixed during the joint estimation.

\subsubsection{Location Prediction with \systemname}
For location prediction task, we predict the next location from the previously seen locations in the training set at $M (\# communities)= 10$ and at various top-$K$ ranked (eq.~\ref{eqn:ll}) cutoffs. Since, DH is for clustering event streams, therefore we make use of STHP and RMTPP as baselines for the Location Prediction task. Table~\ref{tab:loc_pred_A} shows that \systemname is able to offer significant improvements (18-37\% in the top-$5$). As we increase $K$, these gains diminish because the number of candidate locations saturates. We also study the effect of $A_{ij}$(influence matrix) and $\mu$ (base intenisty) over \systemname's performance. It can be observed that without $A_{ij}$, the \systemname's performance degrades signifying \systemname's ability to capture the underlying diffusion process well. 

	\begin{table}[tb]
		\centering
		\caption{Comparison of \systemname with other baselines and \systemname without $A_{ij}$ and $\mu$}
		\label{tab:loc_pred_A}
		\begin{tabular}{ l l l l l l l }
			\toprule
			Dataset & {K} & STHP & RMTPP &\systemname w/o Aij &\systemname w/o $\mu$ & \systemname \\
			\midrule
			\multirow{4}{1.2cm}{SA (\#testcases = 2805)} & 5 & 287 & 250 & 279 & 331 &\textbf{339} \\
			&10 & 455 & 499 & 478 & 589 &\textbf{593} \\
			& 20 & 950 & 951 & 911 & 1038 &\textbf{1043} \\
			&	50 & 1539 & 1539 & 1520 & 1664 &\textbf{1666} \\
			\midrule
			\multirow{4}{1.2cm}{US (\#testcases = 4395)}& 5 & 153 & 172 & 172 & 231 & \textbf{237} \\
			& 10 & 456 & 467 & 445 & 507 & \textbf{512} \\
			& 20 & 870& 888 & 919 & 920 & \textbf{927} \\
			& 50 & 1700& 1691 & \textbf{1759}& 1668 & 1673\\
			\bottomrule
		\end{tabular}

	\end{table}
\subsubsection{Impact of \#Communities}
\begin{figure}[t]
	\centering
	\begin{subfigure}[b]{0.33\textwidth}
		\includegraphics[width=\linewidth]{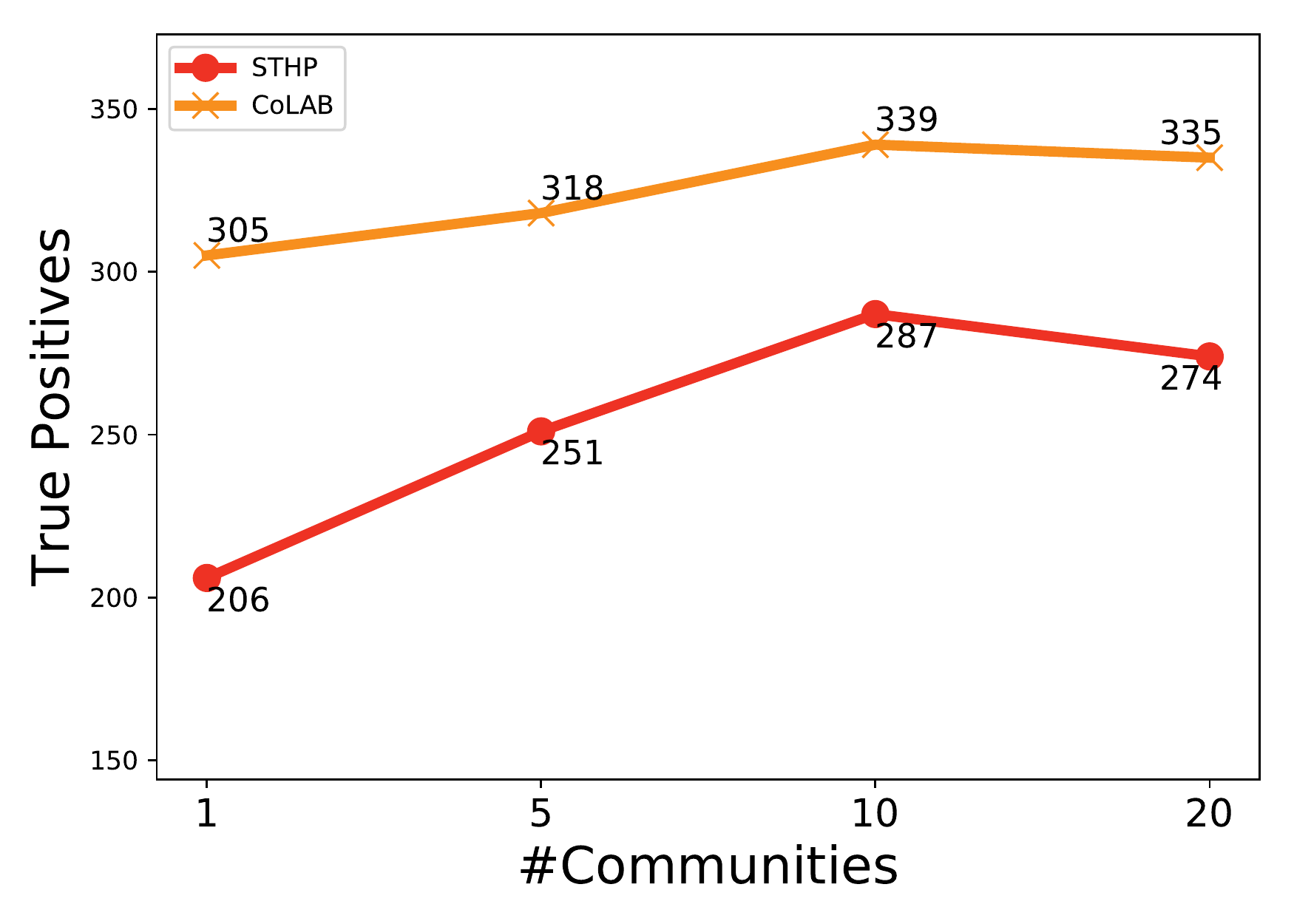}
		\caption{K=5}
	\end{subfigure}%
	\begin{subfigure}[b]{0.33\textwidth}
		\includegraphics[width=\linewidth]{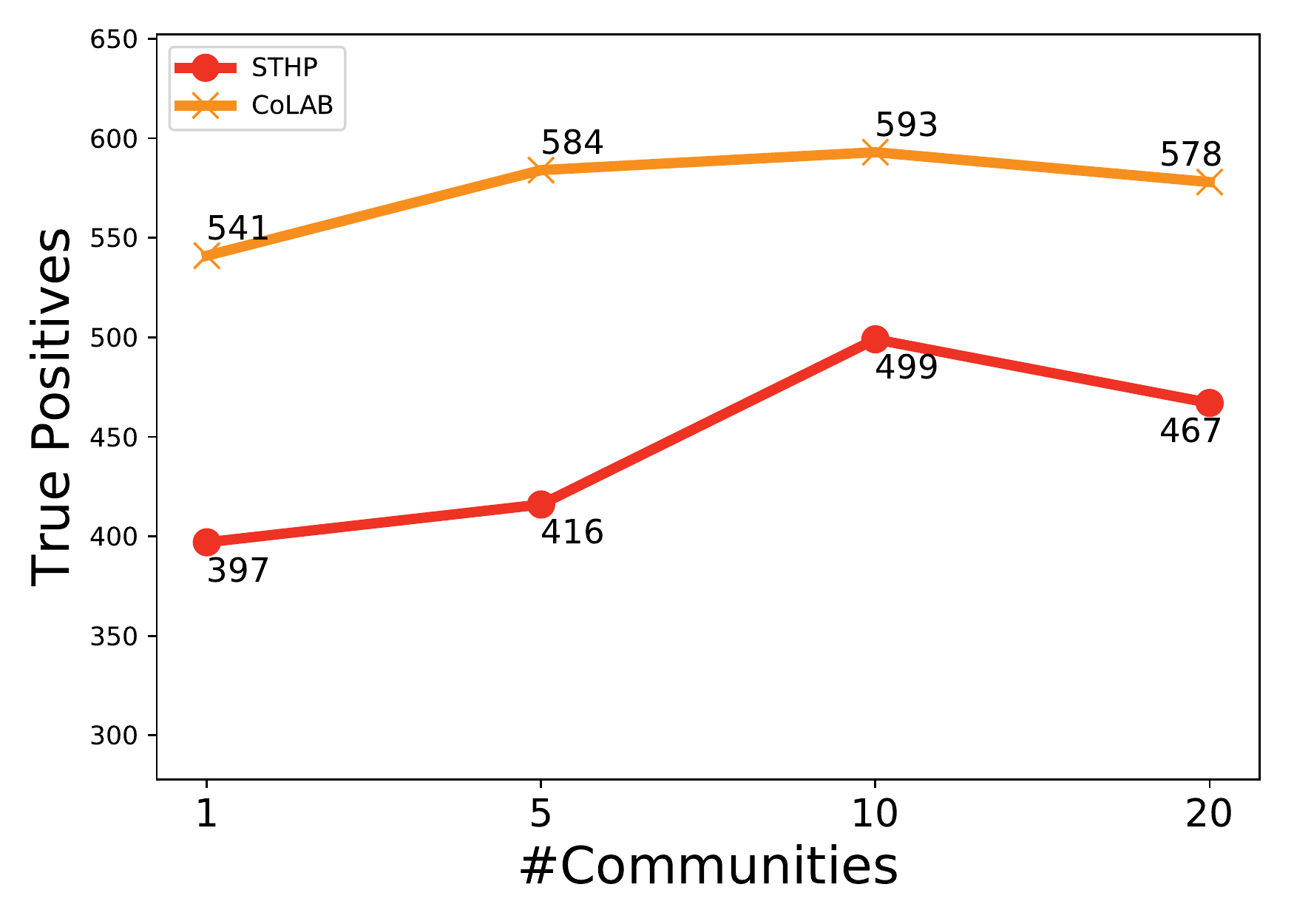}
		\caption{K=10}
	\end{subfigure}%
	\begin{subfigure}[b]{0.33\textwidth}
		\includegraphics[width=\linewidth]{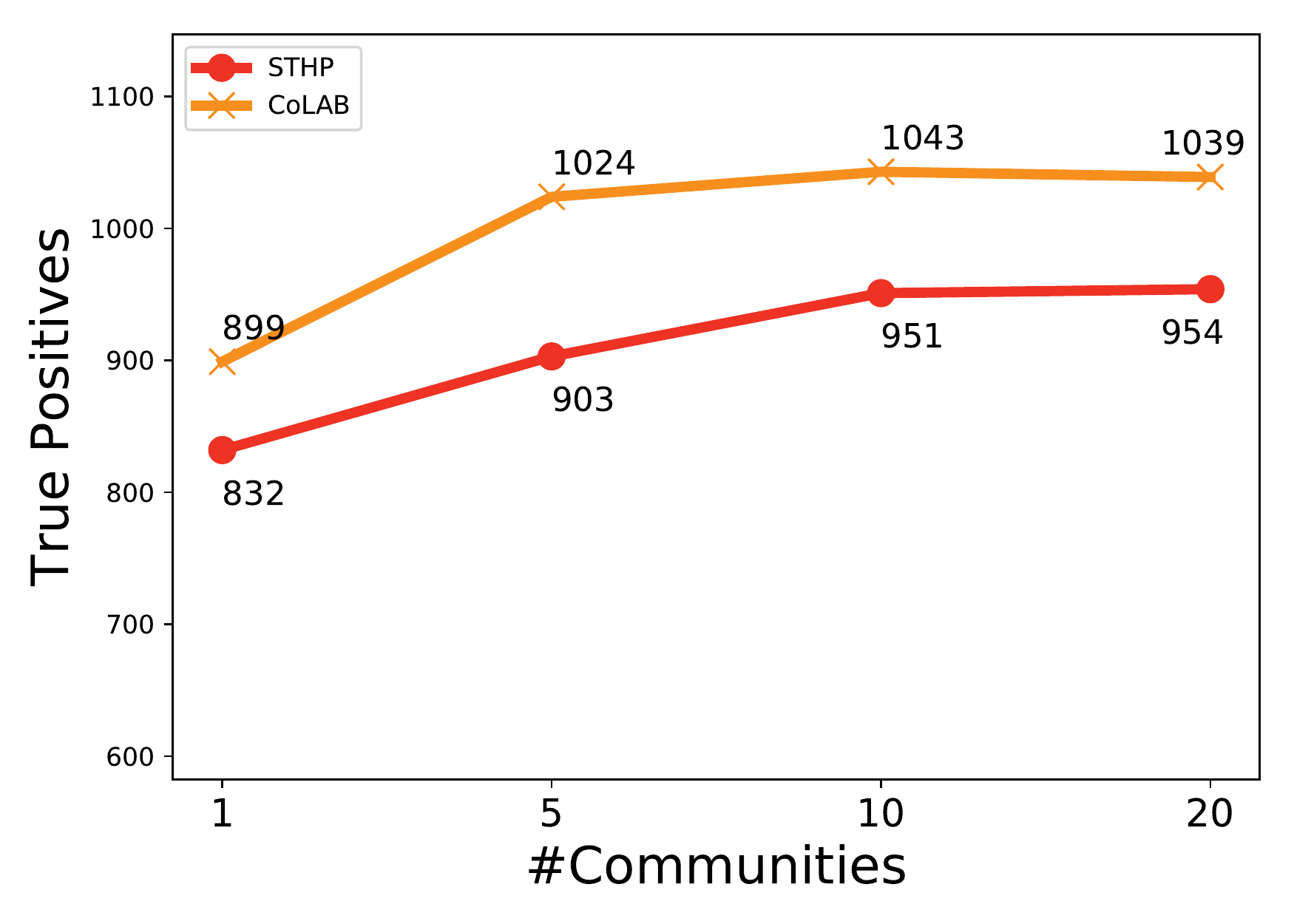}
		\caption{K=20}
	\end{subfigure}  
	\caption{Location Prediction Results for Varying M over SA}
	\label{fig:loc_M_SA}
\end{figure}

\begin{figure}[t]
	\centering
	\begin{subfigure}[b]{0.33\textwidth}
		\includegraphics[width=\linewidth]{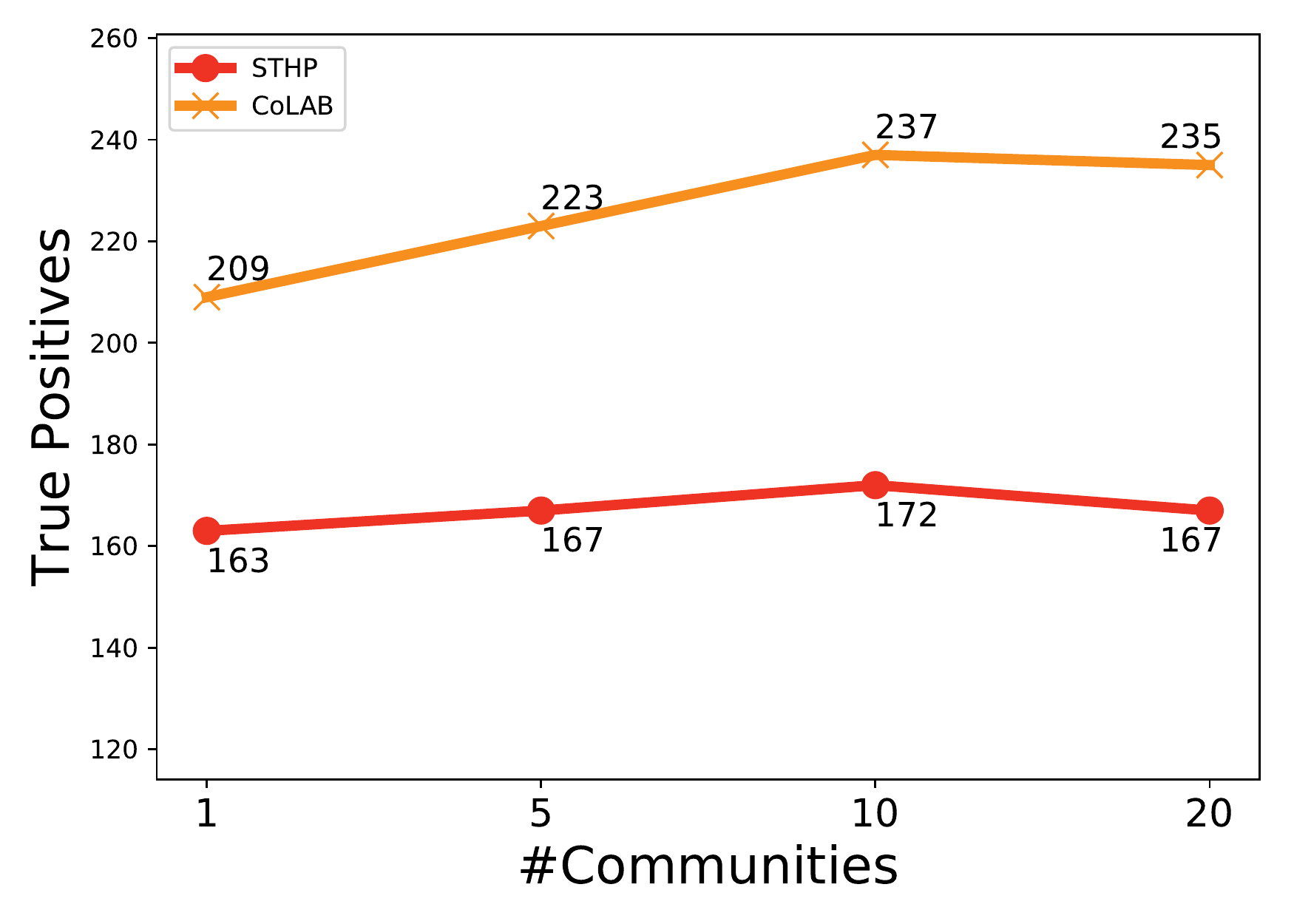}
		\caption{K=5}
	\end{subfigure}%
	\begin{subfigure}[b]{0.33\textwidth}
		\includegraphics[width=\linewidth]{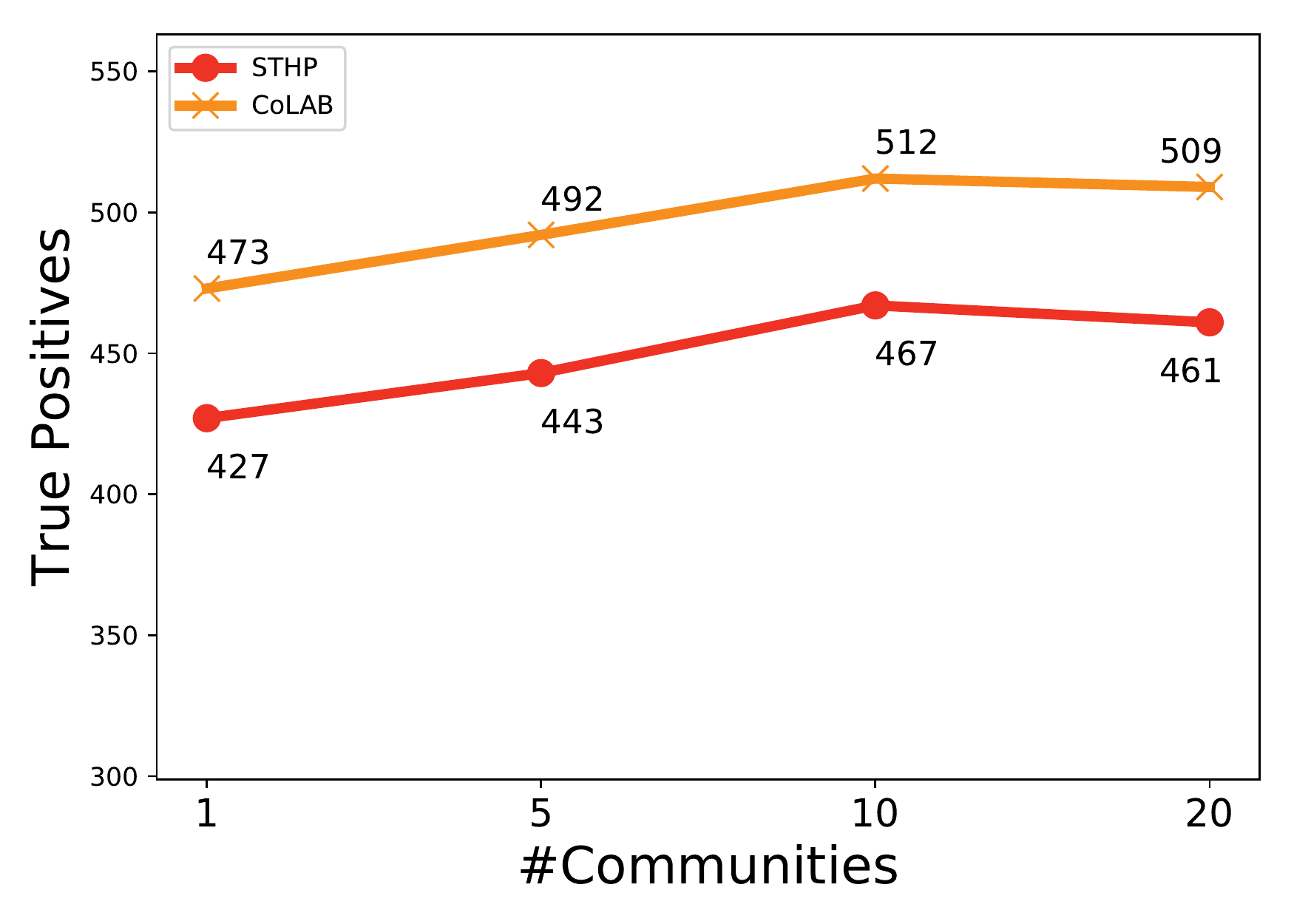}
		\caption{K=10}
	\end{subfigure}%
	\begin{subfigure}[b]{0.33\textwidth}
		\includegraphics[width=\linewidth]{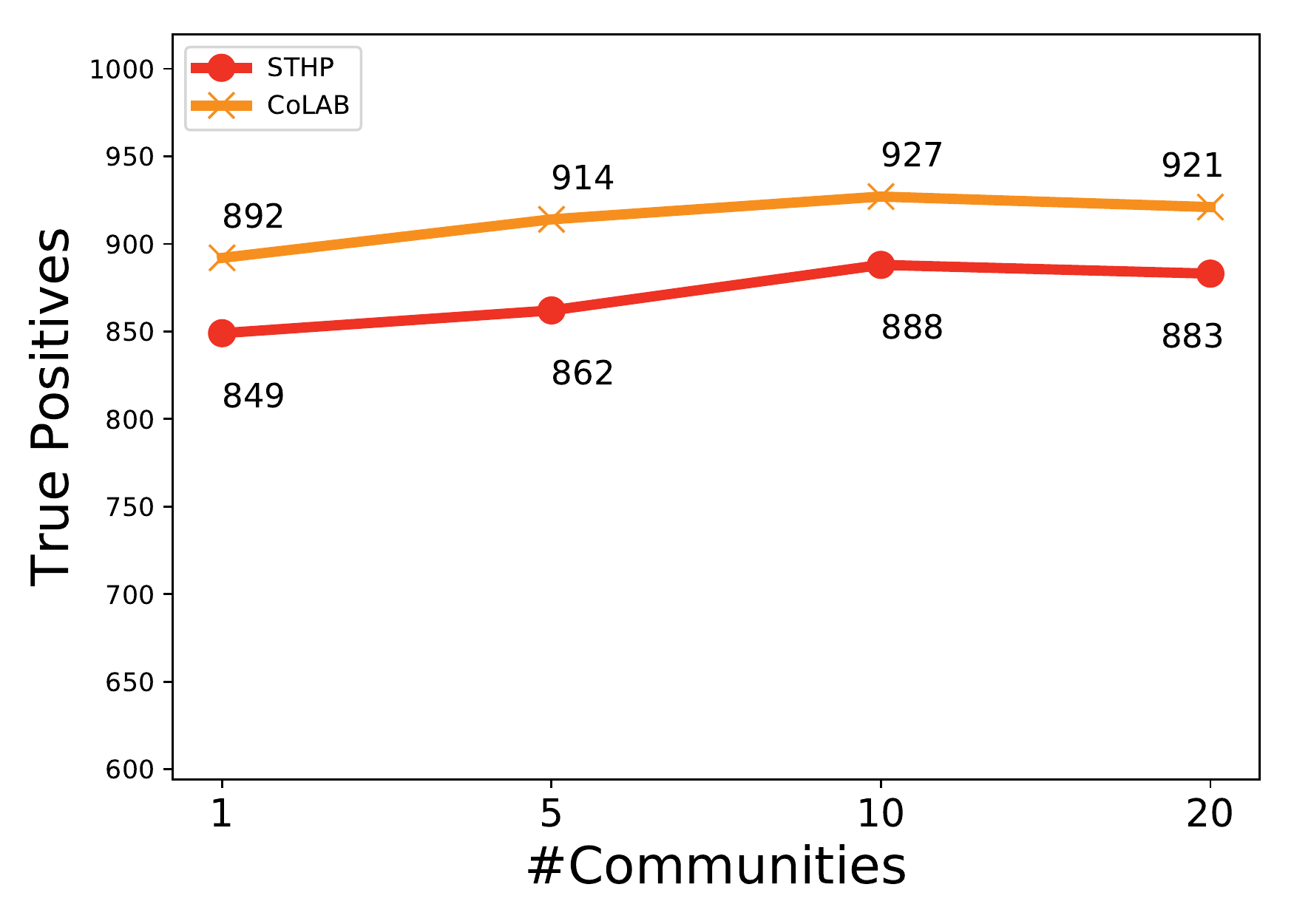}
		\caption{K=20}
	\end{subfigure}
	\caption{Location Prediction Results for Varying M over US }
	\label{fig:loc_M_US}
\end{figure}

In figures~\ref{fig:loc_M_SA} and \ref{fig:loc_M_US} we study the impact of $M$, over SA and US data we observe that with increasing $M$ the prediction accuracy improves and then diminishes, signifying optimal value of $M$ for better predictions. \systemname performs significantly better than STHP, primarily due to the better estimation of $A_{ij}$ (influence matrix), because of the presence of category information in the \systemname model. 

\subsection{Community Assessment}
\begin{figure}[h!]
	\centering
	\begin{subfigure}[b]{0.3\textwidth}
		\frame{\includegraphics[height=3cm]{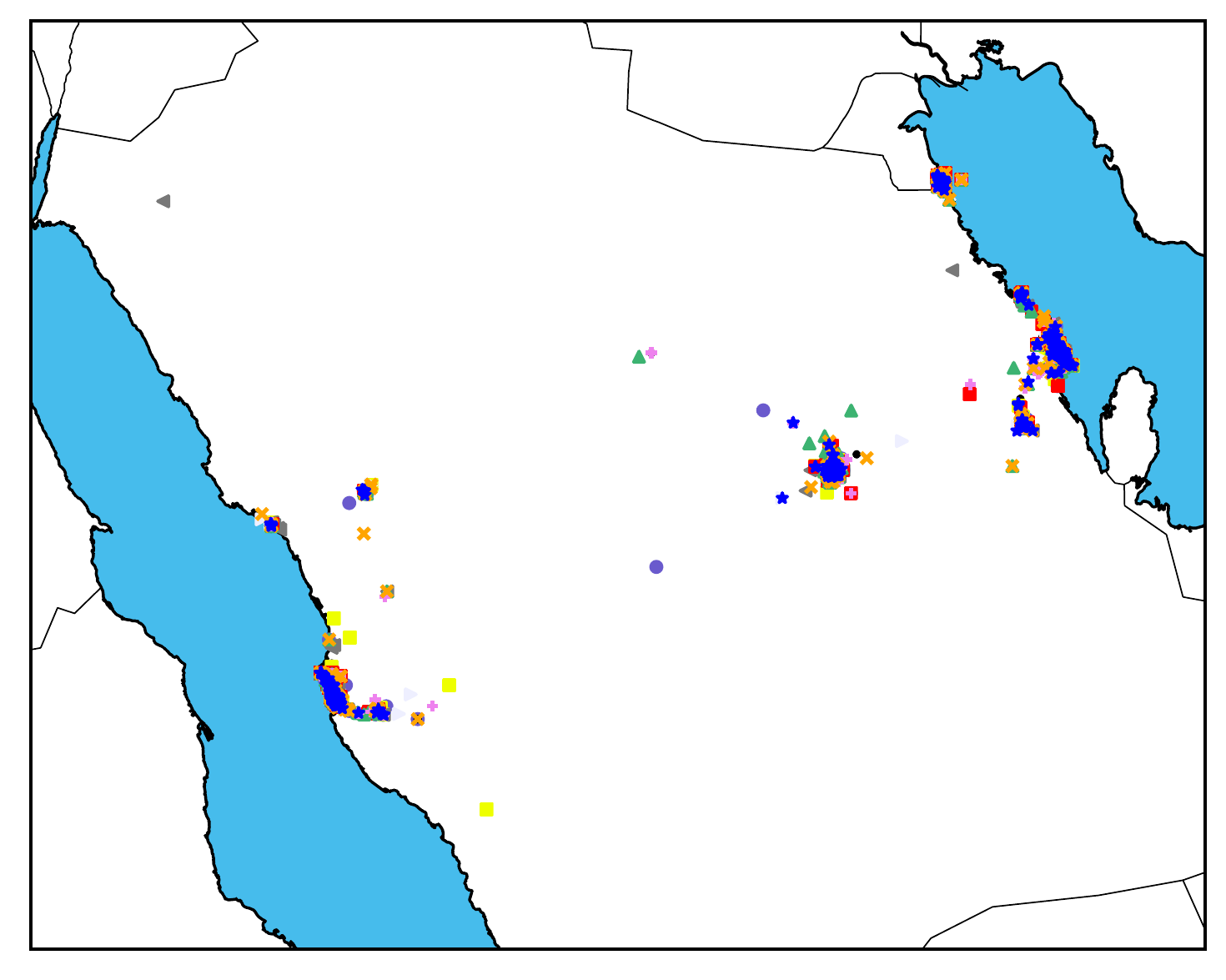}}
		\caption{SA}
	\end{subfigure} \quad
	\begin{subfigure}[b]{0.4\textwidth}
		\frame{\includegraphics[height=3cm]{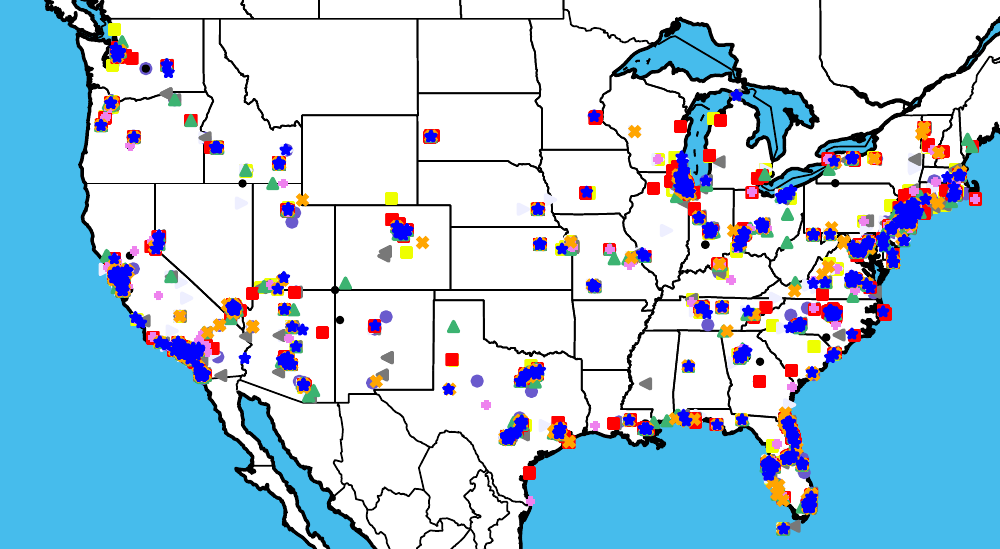}}
		\caption{US}
	\end{subfigure}
	\caption{Spatio-Temporal Activity-driven Latent Communities Captured by \systemname}
	\label{fig:acc2}
\end{figure}

We plot communities over SA and US data in Figure~\ref{fig:acc2}, where colored dots represents a check-in. In figure ~\ref{fig:acc2}, it can be observed that; (i) \textit{Overlap} between communities due to data concentration in cities. (ii) \systemname is able to capture communities across cities. 

Unfortunately, we lack the community ground truth for users, making communities assessment a non-trivial task. Thus, we use a metric a joint loss function for the intra-community properties through a mixture of (i)~\emph{category loss} ($\mathcal{L}_{cat}$) and (ii)~\emph{location loss} ($\mathcal{L}_{loc}$).
\noindent
\textbf{Category Loss:} We consider all categories associated to locations as independent \textit{marks} of a point process. Hence to estimate the category affinity in a community, we consider similarity among the check-in categories using pre-trained word embeddings~\cite{Mikolov:2018} and devise a \textit{loss} function.
\begin{equation}
\mathcal{L}_{cat} = \frac{1}{|\mathbf{T}|}  \sum_{E_n \in \mathbf{T}} \sum_{g \in \mathcal{M}} \left\{1 -  \frac{\mathbf{v}_{E_n} \cdot \mu_g}{||\mathbf{v}_{E_n}||_{2}||\mu_g||_{2}}\right\} \cdot \Phi(E_n, g)
\end{equation}
where $\mathbf{T}$ represents test data, $\mu_g$ is \emph{category mean} for a community $g$ using $K_{cat}$ frequent categories, $\mathbf{v}_{E_n}$ is the category vector for event $E_n$; $\Phi(E_n, g)$ indicates whether $E_n$ is assigned to community $g$, with $\mathcal{M}$ as all communities. Table \ref{tab:cal_loss_all} demonstrates \systemname's ability to capture category dynamics across communities. Although, it can be observed that at $K_{cat}$ = 10 Dirichlet Hawkes performs better because with most frequent categories like restaurant, coffee shop etc. DH assigns it most of the communities. Note that, DH is unable to capture communities with varied categories as seen at $K_{cat}$ = 50 and 100, that \systemname performs better.
\\
\textbf{Location Loss:}\cite{Cho:2011:FMU:2020408.2020579} show that users tend to visit nearby locations given we ignore the bias of loyalty. Hence ideally, a community of users not spatially dispersed in their checkin characteristics should be distinct from another community with checkins spanning large distances. We capture this through a distance based \textit{k-means} loss ($\mathcal{L}_{loc}$) with cluster means($\mu_{l}$) for checkin coordinates for each community. In table \ref{tab:loc_loss_all} we can see \systemname performs significantly better than other baselines because \systemname can better capture the geographical dispersion in communities.

\begin{table}
\centering
	\caption{Results for Category Loss}
	\label{tab:cal_loss_all}
	\begin{tabular}{  l l p{1.75cm} p{1.75cm} p{1.75cm} p{1.75cm} p{1.75cm} }
		\toprule
		Dataset & $K_{cat}$ & \multicolumn{2}{ l }{Sequence Mining} & DH & STHP & \systemname  \\
		\cmidrule(lr){3-4}
		& & Daily & Weekly & & \\
		\midrule
		 \multirow{3}{*}{SA}& 10 & 250.24 & 236.19 & \textbf{103.47} & 125.17 & 119.41 \\
		&50 & 1118.23 & 1089.27 & 862.05 & 842.63 & \textbf{826.72} \\
		& 100 & 2007.93 & 2120.76 & 1983.61 & 1784.04 & \textbf{1749.37} \\
		\midrule
		\multirow{3}{*}{US} & 10 & 248.45 & 217.56 & \textbf{98.37} & 118.32 & 113.86 \\
		& 50 & 956.87 & 990.45 & 781.83 & 793.07 & \textbf{771.15} \\
		& 100 & 1907.84 & 2020.49 & 1901.37 & 1605.64 & \textbf{1583.02} \\
		\bottomrule
	\end{tabular}
\end{table}

\begin{table}
\centering
\caption{Results for Location Loss}
\label{tab:loc_loss_all}
	\begin{tabular}{ l l  p{1.75cm} p{1.75cm} p{1.75cm} p{1.75cm}}
		\toprule
		\multirow{2}{*}{Datasets} & \multicolumn{2}{l}{Sequence Mining}& DH &STHP & \systemname \\
		\cmidrule(lr){2-3}
		& Daily & Weekly & & & \\
		\midrule
		SA & 600.67 & 547.56 & 413.16 & 306.73 & \textbf{298.09} \\
		US & 1127.34 & 1067.50 & 1039.08 & 849.92 & \textbf{834.64} \\
		\bottomrule
	\end{tabular}

\end{table}
\subsubsection{Qualitative Assessment}
\begin{figure}[htb]
\centering
\begin{subfigure}[b]{0.35\columnwidth}
\includegraphics[width=\linewidth]{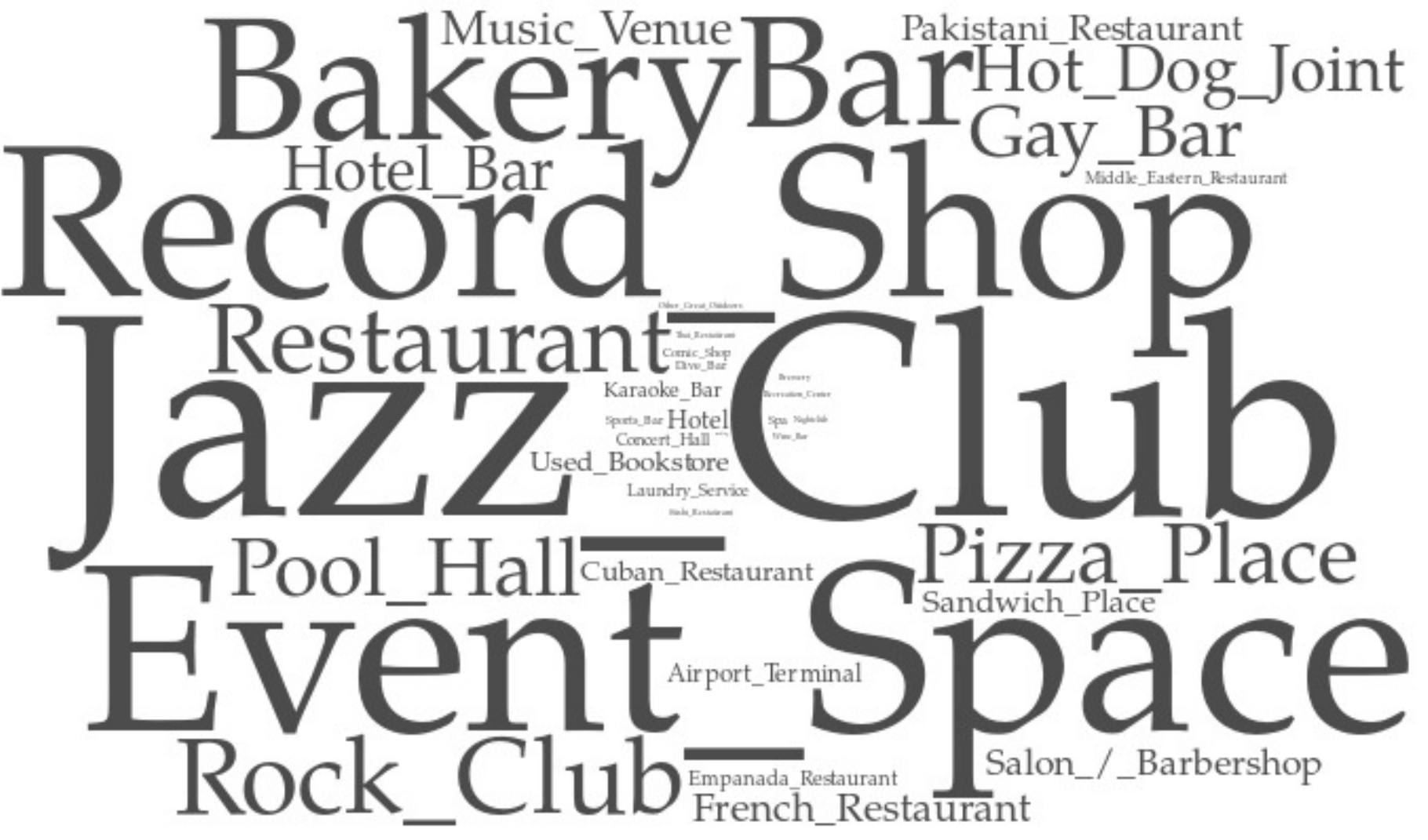}
\caption{Community A}
\end{subfigure}
\begin{subfigure}[b]{0.35\columnwidth}
\includegraphics[width=\linewidth]{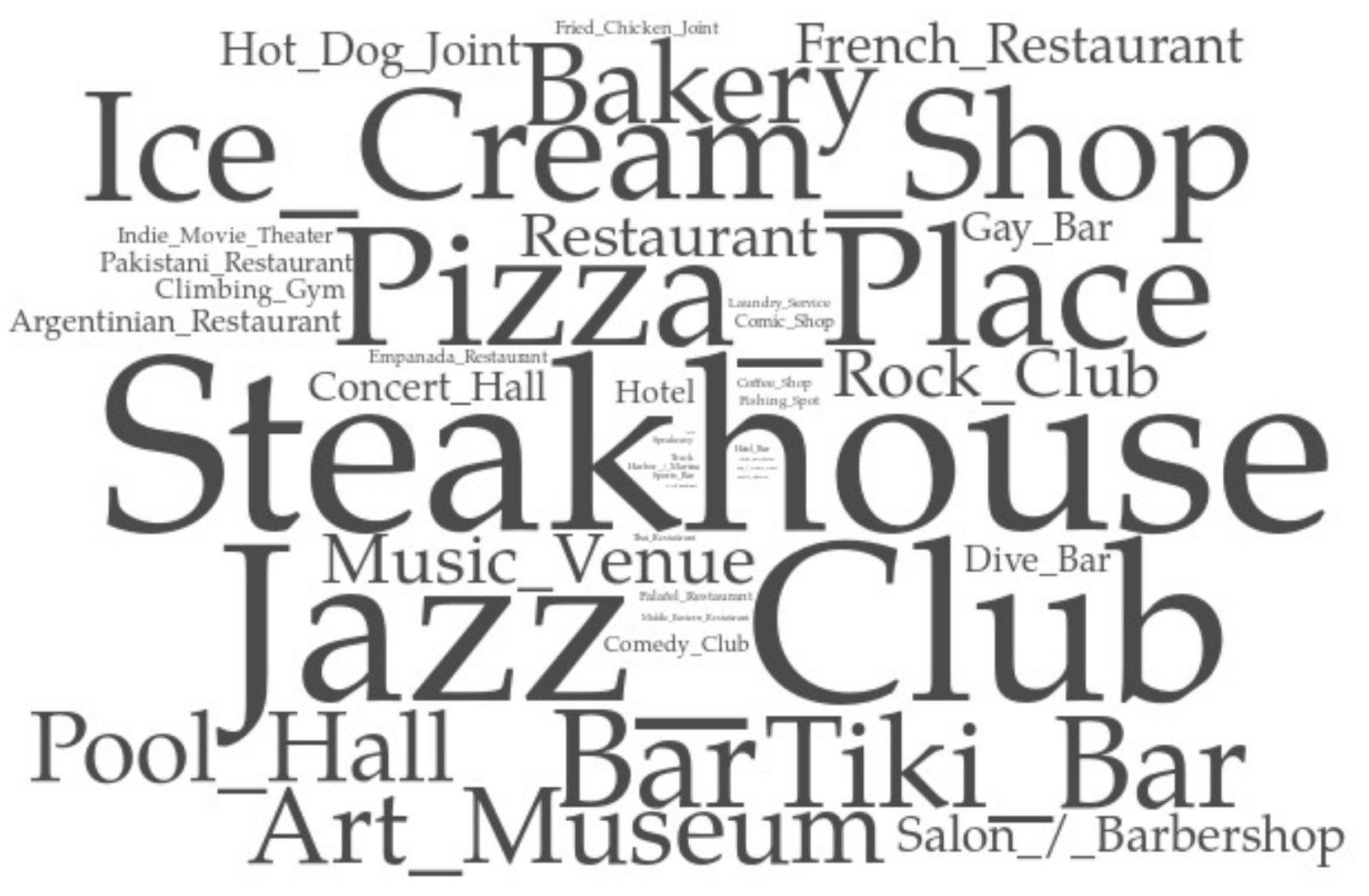}
\caption{Community B}
\end{subfigure}
\caption{Word Cloud of Categories in Two Communities from US}
\label{fig:wordcloud}
\end{figure}

We claim that a user in a community will display an affinity towards certain categories. For US data, figure~\ref{fig:wordcloud} shows even with highly overlapping venue categories, our model finds the intricate differences between a community with affinity to music (a) and a community with affinity towards food/bar joints. The word clouds for SA dataset shows similar properties and have been avoided for brevity.
\section{Conclusion}
In this paper, we presented \systemname that uses spatio-temporal Hawkes process to infer the implicit communities using a novel stochastic variational inference technique. Empirical evaluations over synthetic as well as real-world datasets highlight its prowess with significant improvements in location and community detection tasks. This illustrates the effectiveness of the modelling used in CoLAB in generating user communities even in the absence of social connectedness information. In future work, we would like to explore scalability of \systemname through sample-based inference techniques.

 \bibliographystyle{splncs04}
\bibliography{bibfile}	
\end{document}